\documentclass[twoside,12pt]{article}
\usepackage{epsfig}

\newcommand{\be}{\begin{equation}}
\newcommand{\ee}{\end{equation}}
\newcommand{\bea}{\begin{eqnarray}}
\newcommand{\eea}{\end{eqnarray}}

\newcommand{\vlk}{$V_{\rm low-k}$ }
\newcommand{\vlkn}{$V_{\rm low-k}$}
\begin{document}

\title{ \vspace{1cm} Chiral nuclear dynamics with three-body forces}
\author{J.\ W.\ Holt, N.\ Kaiser and W.\ Weise \\
Technische Universit\"at M\"unchen, Germany}
\maketitle
\begin{abstract} 
We review recent progress in implementing high-precision chiral two- and 
three-body forces in nuclear many-body systems beyond light nuclei. We begin
with applications to finite nuclei, which we study through the nuclear shell 
model and self-consistent mean field theory. We then turn our attention to 
infinite nuclear matter treated within the framework of Landau's theory of 
normal Fermi liquids.
\end{abstract}
\section{Introduction}

In recent years, the application of chiral effective field theory methods 
to low-energy nuclear dynamics has resulted in the development of 
high-precision two- and many-body forces consistent with the symmetry 
constraints imposed by the fundamental theory of strong interactions, quantum 
chromodynamics. For recent reviews on this subject, see refs.\ 
\cite{epelbaum06,entem11}. Although such interactions have been highly 
successful in {\it ab initio} calculations of light-nuclei scattering observables, 
binding energies and spectra \cite{epelbaum02,witala06,navratil07}, 
there remains much work to be 
done to implement three-nucleon forces in studies of medium-mass and heavy
nuclei. In this contribution we discuss our recent work to understand the
properties of finite nuclei and infinite nuclear matter from many-body 
perturbation theory employing realistic N$^3$LO chiral two-nucleon forces 
supplemented with the leading-order N$^2$LO three-body force. For this 
purpose it is convenient to employ as well low-momentum nucleon-nucleon (NN) 
interactions \cite{vlowk,vlowkreview}, which exhibit favorable convergence 
properties in perturbative calculations \cite{bogner05}.

\section{In-medium effective nucleon-nucleon interactions \label{imenni}}

To facilitate the implementaion of three-body forces in nuclear many-body calculations, 
it is useful to construct medium-dependent NN interactions
that reflect the most important physical features of the underlying three-nucleon
force. Here we focus on a background medium of isospin-symmetric nuclear
matter and compute the density-dependent NN interaction to one-loop order from the
N$^2$LO chiral three-nucleon interaction. In fact, the effective interaction in 
nuclear matter with isospin asymmetries up to $\delta_{np}= (\rho_n-\rho_p)/\rho 
\simeq 0.2$ is already well-approximated by the effective interaction for
$\delta_{np} = 0$ \cite{holt10}. For detailed discussions regarding the
in-medium interaction in pure neutron matter, we refer the reader to refs.\ 
\cite{holt10,hebeler10}.

The leading-order chiral three-nucleon force includes three separate
components of varying range. The two-pion exchange contribution 
$V_{3N}^{(2\pi)}$ consists of terms proportional to the low-energy constants 
$c_{1,3,4}$, which arise already in the two-nucleon interaction and can
therefore be constrained by fits to NN scattering phase shifts. The one-pion exchange component
$V_{3N}^{(1\pi)}$ is proportional to the low-energy constant $c_D$, while
the short-range contact contribution $V_{3N}^{(ct)}$ introduces the
low-energy constant $c_E$. Both $c_D$ and $c_E$ are commonly fit to the 
properties of $A=3,4$ systems. In the present work we use two different parameterizations 
of the low-energy constants $c_E, c_D$ and $c_{1,3,4}$:
\be
c_E = -0.205, \hspace{.1in} c_D = -0.20, \hspace{.1in} c_1 =-0.81\, {\rm GeV}^{-1}, \hspace{.1in} 
c_3=-3.2\,{\rm GeV}^{-1}, \hspace{.1in} c_4 =5.4 \, {\rm GeV}^{-1},   
\label{lecn3lo}
\ee
\be
c_E = -0.625, \hspace{.1in} c_D = -2.06, \hspace{.1in}  c_1 =-0.76\, {\rm GeV}^{-1}, \hspace{.1in} 
c_3=-4.78\,{\rm GeV}^{-1}, \hspace{.1in} c_4 =3.96 \, {\rm GeV}^{-1}, 
\label{lecvlk}
\ee 
which are consistent with the Idaho N$^3$LO chiral NN interaction \cite{gazit09} and the
low-momentum NN interaction \vlk evolved to a decimation scale of $\Lambda = 2.1$\,fm$^{-1}$ 
\cite{bogner05}, respectively.
\begin{figure}[tb]
\begin{center}
\begin{minipage}[t]{18.5 cm}
\centering
\epsfig{file=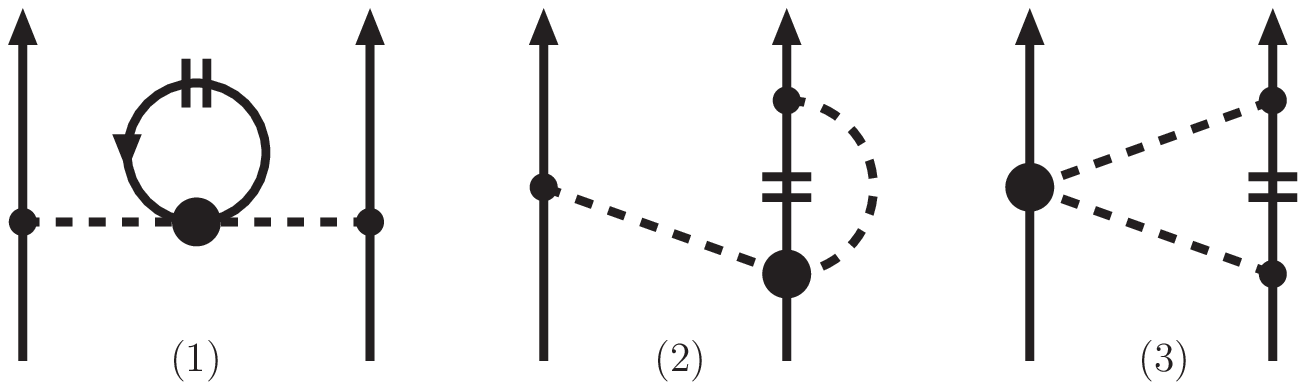,scale=.59} \hspace{.3in}
\epsfig{file=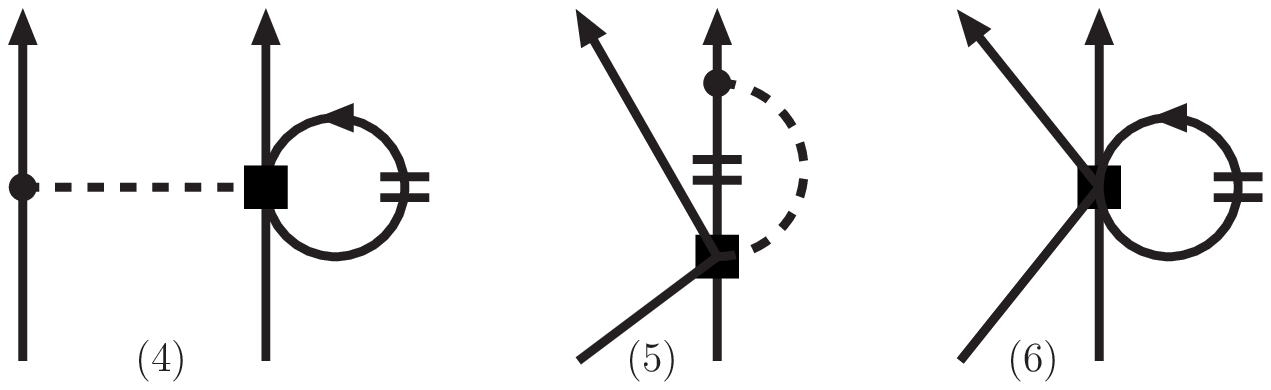,scale=.59}
\caption{In-medium NN interaction generated by the three different components
of the chiral three-nucleon force. The short double-line symbolizes the filled 
Fermi sea of nucleons, and reflected diagrams are not shown. \label{ddnnint}}
\end{minipage}
\end{center}
\end{figure}

At one-loop order there are six distinct diagrams, shown in 
Fig.\ \ref{ddnnint}, contributing to the in-medium NN interaction $V_{NN}^{\rm med}$. The 
double-line symbolizes summation over the filled Fermi sea of nucleons, or
equivalently, the medium insertion $-2\pi \delta(k_0)\,
\theta(k_f- |\vec k\,|)$ in the in-medium nucleon propagator.
For a background medium at rest, the in-medium scattering amplitude for two 
nucleons in their center of mass frame, $N_1(\vec p\,)+ N_2(-\vec p\,) \to 
N_1(\vec p+\vec q\,) + N_2(-\vec p-\vec q\,)$, has the same general form as 
that given in free-space NN scattering. For a background medium boosted with 
respect to the center of mass frame, the effective interaction 
will in general depend also on the boost momentum.

\begin{figure}[tb]
\begin{center}
\begin{minipage}[t]{0.45\textwidth}
\centering
\includegraphics[scale=.32,angle=270]{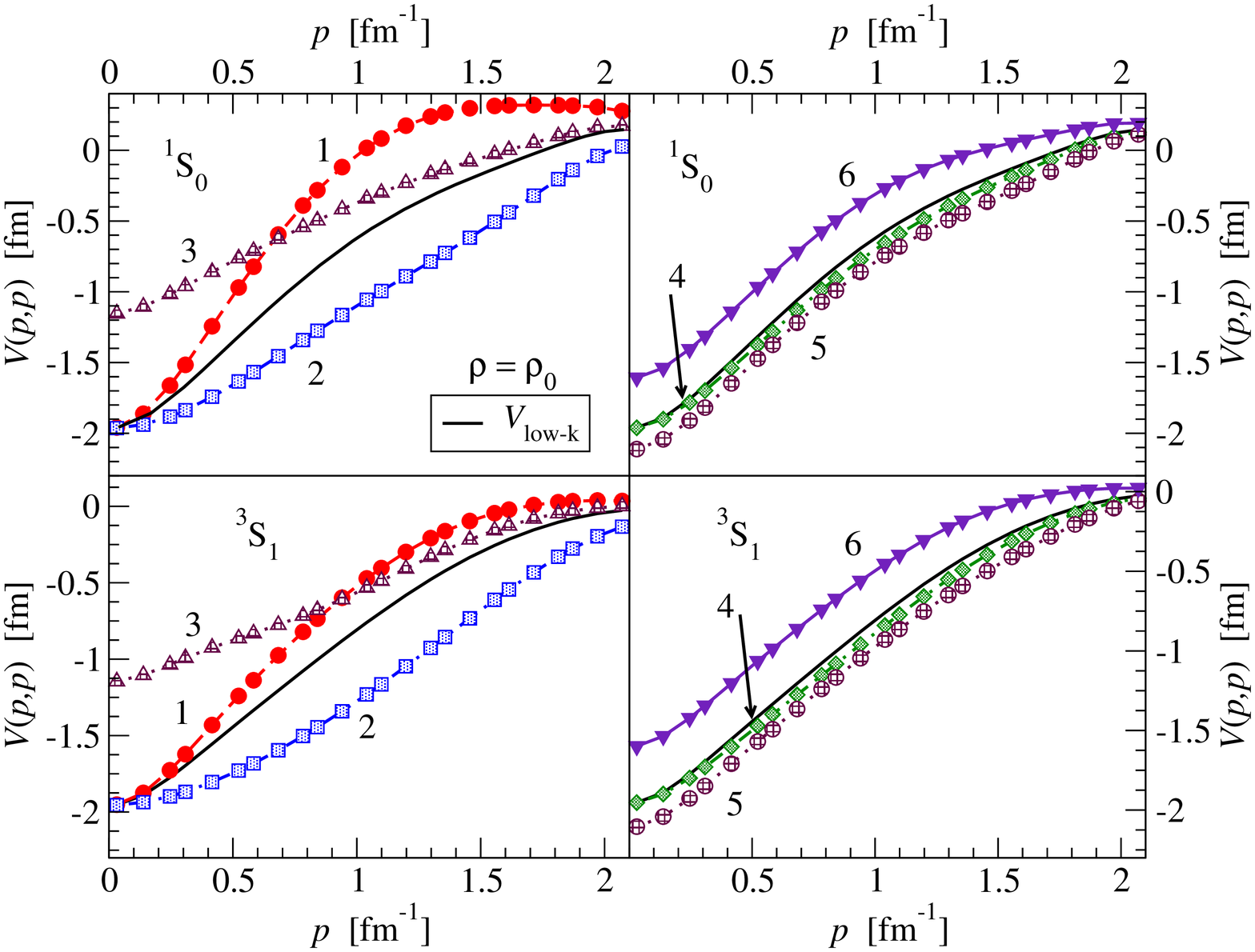}
\end{minipage}
\hspace{.3in}
\begin{minipage}[t]{0.45\textwidth}
\centering
\vspace{40pt}
\includegraphics[scale=.3,angle=270]{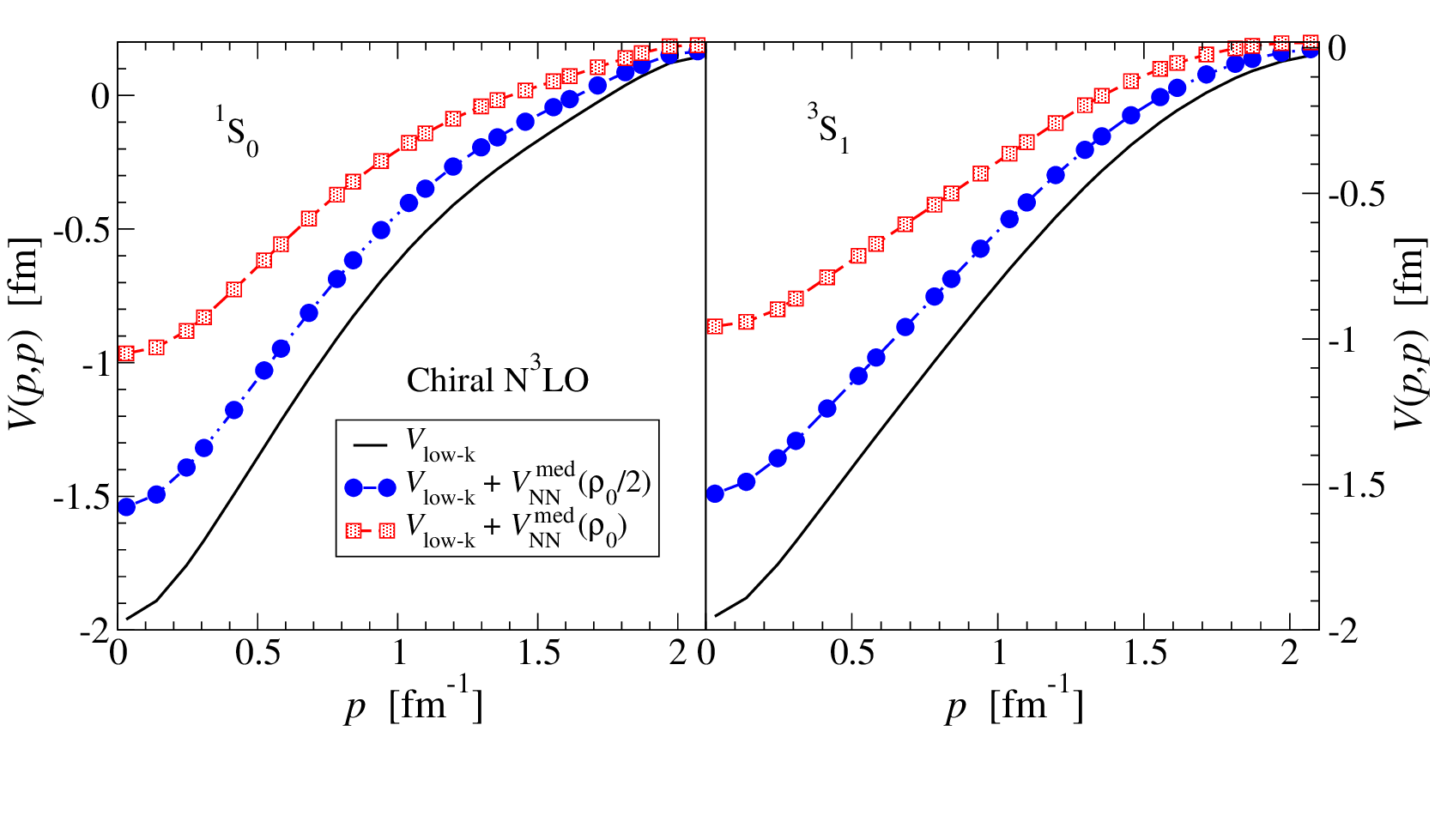}
\end{minipage}
\caption{Left figure: contributions to the $S$-wave effective interaction in symmetric 
nuclear matter (at saturation density, $\rho_0=0.16$ fm$^{-3}$) resulting
from the six diagrams shown in Fig.\ \ref{ddnnint}. The solid black line
shows the matrix elements of the free-space NN potential \vlkn, and the labels 1--6 denote
the contributions from
the corresponding diagrams in Fig.\ \ref{ddnnint}. Right figure:
contributions from all components of the in-medium NN interaction to the $L=0$ partial
waves at $\rho = \rho_0, \rho_0/2$.
\label{n3lopw}}
\end{center}
\end{figure}

In Fig.\ \ref{n3lopw} we plot the $^1S_0$ and $^3S_1$ diagonal partial-wave matrix 
elements of the low-momentum NN potential \vlk and the modifications resulting from the six
components of $V_{NN}^{\rm med}$ evaluated at nuclear matter saturation density ($\rho_0 = 0.16$\,fm$^{-3}$) 
with low-energy constants given in 
eq.\ (\ref{lecvlk}). In general, the three contributions from $V_{3N}^{(2\pi)}$
can be a significant fraction of the free-space matrix elements of \vlk and are
considerably stronger than those arising from the one-pion exchange and contact
three-nucleon forces. The pion self-energy correction (diagram (1) in
Fig.\ \ref{ddnnint}) gives rise to an enhancement of the bare $1\pi$-exchange, which
can be interpreted in terms of a reduced pion decay constant, 
$f_{\pi,s}^{*2}=f_\pi^2+2c_3\rho$. In contrast, the vertex correction (diagram 
(2) in Fig.\ \ref{ddnnint}) from $V_{3N}^{(2\pi)}$ reduces one-pion exchange, 
an effect that can be approximated by introducing a reduced nucleon axial-vector constant
$g_A^*$. Taken together, these two terms largely cancel in all partial waves,
which leaves the Pauli-blocked two-pion exchange process (diagram (3) in Fig.\ 
\ref{ddnnint}) as a dominant contribution. The contributions from the one-pion
exchange three-body force $V_{3N}^{(1\pi)}$ are quite small. Since $c_D$ is 
negative, the vertex correction (diagram (4) in Fig.\ \ref{ddnnint}) reduces 
the bare $1\pi$-exchange, but only by about $16\%$ at normal nuclear matter density.
The pion-loop correction to the NN contact interaction (diagram (5) in
Fig.\ \ref{ddnnint}) acts only in the two relative $S$-waves as well as the 
$S-D$ mixing matrix element. In relative $S$-waves, it reduces the repulsive 
contribution from the pure contact term (diagram (6) in Fig.\ \ref{ddnnint})
by about one-half. In the right plot of Fig.\ \ref{n3lopw} we sum all
six contributions, which together give rise to significant additional repulsion 
that increases with the density and which provides a mechanism for nuclear
matter saturation when employing low-momentum potentials \cite{bogner05}.

The resulting in-medium effective interaction has been employed already in shell model
studies of finite nuclei, where it was found to play a key role in 
enhancing the $\beta$-decay lifetime of $^{14}$C \cite{holt09} toward its 
known achaeologically-long value of 5730 years. An alternative approach for
finite nuclei calculations is to evaluate the three-nucleon force in a shell model 
basis and subsequently sum over 
the filled orbitals in the closed nuclear core. Such a method has been used to
explore the effects of three-nucleon forces in neutron-rich oxygen and calcium 
isotopes \cite{jdholt11}.

\section{Microscopic energy density functionals}

To study the binding energies and charge radii of nuclei across the periodic
chart, the most successful theoretical framework has been self-consistent
mean field theory based on phenomenological nuclear energy density functionals \cite{bender03}. 
A microscopic foundation for such functionals is provided within the framework of 
many-body perturbation theory by the density matrix expansion of Negele and Vautherin 
\cite{negele72}, in which the highly nonlocal expression for the energy is 
expanded in terms of local densities and currents (and their gradients) resulting in
 a generalized Skyrme functional with density-dependent couplings. 
\begin{figure}[tb]
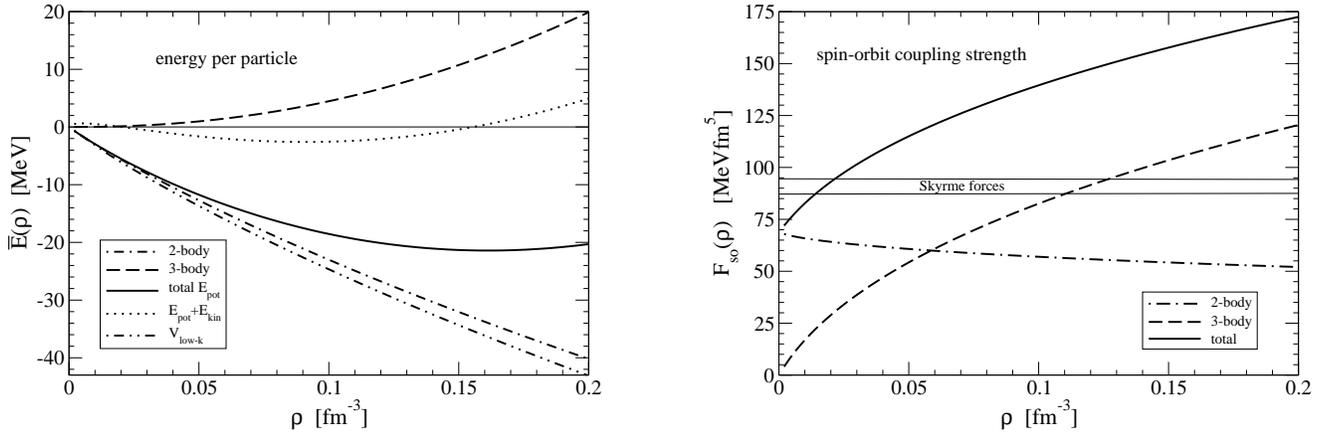

\begin{minipage}[t]{9.25 cm}
\centering
\epsfig{file=ebar23.eps,scale=.32}
\end{minipage}
\begin{minipage}[t]{9.25 cm}
\centering
\epsfig{file=fso23.eps,scale=.32}
\end{minipage}
\caption{Contributions to the energy per particle ${\bar E}(\rho)$ of 
isospin-symmetric nuclear matter and to the spin-orbit strength parameter $F_{so}$ 
as a function of the nuclear density $\rho$.   \label{fso23}}
\end{figure}

Recently Gebremariam, Duguet and Bogner \cite{geb10} have developed
an improved density matrix expansion that gives rise to a much improved
description of the spin density matrix. The form of the density matrix they
find (neglecting isovector terms) can be expanded in 
relative and center-of-mass coordinates, $\vec a$ and $\vec r$, as follows:
\begin{eqnarray} 
\sum_{\alpha}\Psi_\alpha( \vec r -\vec a/2)\Psi_\alpha^
\dagger(\vec r +\vec a/2) &=& {3 \rho\over a k_f}\, j_1(a k_f)-{a \over 2k_f} 
\,j_1(a k_f) \bigg[ \tau - {3\over 5} \rho k_f^2 - {1\over 4} \vec \nabla^2 
\rho \bigg] \nonumber \\ && + {3i \over 2a k_f} \,j_1(a k_f)\, \vec \sigma
\cdot (\vec a \times \vec J\,) + \dots\,, \label{idme} 
\end{eqnarray}
where the spherical Bessel function $j_1(x) = (\sin x - x \cos x)/x^2$. The quantities 
appearing on the right hand side of eq.\ (\ref{idme}) are the (local) nucleon density $\rho(\vec r\,) 
=2k_f^3(\vec r\,)/3\pi^2 =  \sum_\alpha \Psi^\dagger_\alpha(\vec r\,) \Psi_\alpha(\vec r\,)$, 
the (local) kinetic energy density $\tau(\vec r\,) =  \sum_\alpha \vec \nabla 
\Psi^\dagger_\alpha (\vec r\,) \cdot \vec \nabla \Psi_\alpha(\vec r\,)$ and the (local) 
spin-orbit density $ \vec J(\vec r\,) = i \sum_\alpha \vec \Psi^\dagger_\alpha(\vec r\,) 
\vec \sigma \times \vec \nabla \Psi_\alpha(\vec r\,)$.

The energy density functional resulting from this improved density matrix expansion 
has been calculated to three-loop order including $1\pi$-exchange, iterated $1\pi$-exchange
and irreducible $2\pi$-exchange with intermediate $\Delta$-isobars in ref.\ \cite{kaiser10} 
and with a microscopic N$^2$LO nucleon-nucleon potential in ref.\ \cite{geb11}. 
In our work \cite{holt11b}, we have employed a microscopic N$^3$LO chiral NN potential supplemented with
the N$^2$LO chiral three-nucleon force. The two-body interaction is comprised 
of long-range one- and two-pion exchange contributions together with a set of contact terms 
contributing up to fourth power in momenta. The potential, referred to as the N$^3$LOW chiral
NN interaction \cite{n3low}, has a sharp momentum-space
cutoff $\Lambda = 2.1$\,fm$^{-1}$ corresponding to the scale at which low-momentum
interactions become universal and strongly perturbative. The low-energy constants
$c_E, c_D$ and $c_{1,3,4}$ of the leading-order chiral three-nucleon interaction 
are those of eq.\ (\ref{lecvlk}).

Up to second order in spatial gradients, the 
energy density functional relevant for $N=Z$ even-even nuclei reads:
\begin{equation}
{\cal E}[\rho,\tau,\vec J\,] = \rho\,\bar E(\rho)+\bigg[\tau-
{3\over 5} \rho k_f^2\bigg] \bigg[{1\over 2M}-{k_f^2 \over 4M^3}+F_\tau(\rho)
\bigg]  + (\vec \nabla \rho)^2\, F_\nabla(\rho)+  \vec \nabla 
\rho \cdot\vec J\, F_{so}(\rho)+ \vec J\,^2 \, F_J(\rho)\,.
\label{edf}
\end{equation}
Due to the nonlocalities present in the chiral contact interactions, it is 
convenient to compute the density-dependent strength functions 
$\bar E(\rho)$, $F_\tau(\rho)$, $F_d(\rho)$, $F_{so}(\rho)$ and $F_J(\rho)$ 
separately for the finite-range pion-exchange and the zero-range contact components
in the chiral N$^3$LO NN interaction. The leading-order three-nucleon force is 
computed under the assumption that the relevant product of density-matrices can be
represented in momentum space in a factorized form: $\Gamma(\vec p_1,\vec q_1)\,
\Gamma(\vec p_2,\vec q_2)\,\Gamma(\vec p_3,-\vec q_1-\vec q_2)$. 

\begin{figure}[tb]
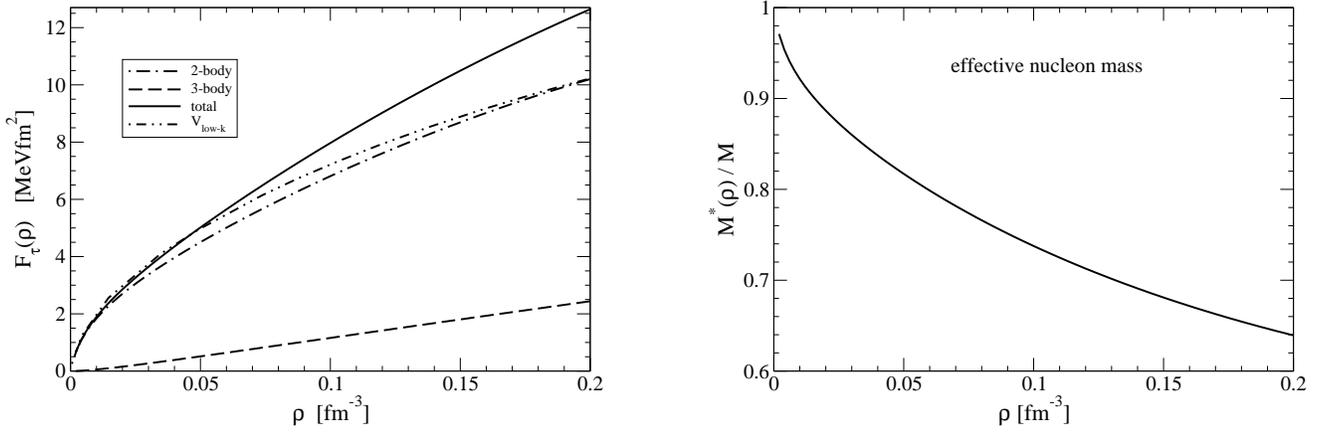

\begin{minipage}[t]{9.25 cm}
\centering
\epsfig{file=ftau23.eps,scale=.32}
\end{minipage}
\begin{minipage}[t]{9.25 cm}
\centering
\epsfig{file=mstar23.eps,scale=.32}
\end{minipage}
\caption{Contributions to the strength parameter $F_\tau$ and effective mass $M^*$ as a function of the 
nuclear density $\rho$. \label{mstar23}}
\end{figure}

\begin{figure}[t]
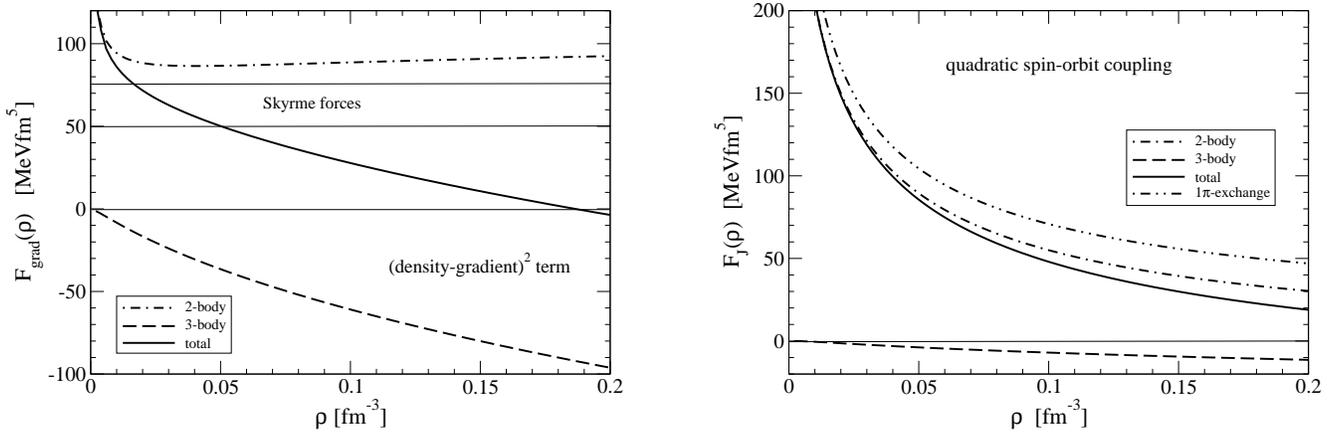

\vspace{.2in}
\begin{minipage}[t]{9.25 cm}
\centering
\epsfig{file=fgrad23.eps,scale=.32}
\end{minipage}
\begin{minipage}[t]{9.25 cm}
\centering
\epsfig{file=fj23.eps,scale=.32}
\end{minipage}
\caption{Contributions to the strength parameters $F_\nabla$ and $F_J$ as a function of the 
nuclear density $\rho$. \label{fj23}}
\end{figure}

In Figs.\ \ref{fso23}--\ref{fj23} we plot the density dependence of the five 
strength functions $\bar E(\rho)$, $F_\tau(\rho)$, $F_d(\rho)$, $F_{so}(\rho)$ and 
$F_J(\rho)$.
At first order in perturbation theory, low-momentum interactions are typically
underbound after including the leading-order chiral three-nucleon force \cite{bogner05}, and this
is reflected in the left plot of Fig.\ \ref{fso23}, where it is seen that both the saturation
density and binding energy are too small. In the right plot of Fig.\ \ref{fso23} we show 
the spin-orbit strength, which receives
a very large positive contribution from the N$^2$LO chiral three-body force that
increases the strength at $\rho = \rho_0/2$ beyond values typical of Skyrme functionals. The 
strength function $F_\tau(\rho)$ (shown in the left plot of Fig.\ \ref{mstar23}) is related
to the effective nucleon mass $M^*$, shown in the right plot of Fig.\ \ref{mstar23}.
At saturation density, the effective nucleon mass is $M^* \simeq 0.7\,M_N$, within the 
range of values used in phenomenological functionals. Finally, we consider 
$F_\nabla (\rho)$ and $F_J(\rho)$ shown in Fig.\ \ref{fj23}. The former encodes
the energy due to density gradients at the Fermi surface, and at half nuclear
matter saturation density it achieves values slightly below those
extracted from Skyrme functionals. The strong variation at low densities 
of $F_J(\rho)$ originates from the dominant $1\pi$-exchange contribution \cite{kaiser10}.
The resulting microscopic energy density functional is an encouraging start and 
leaves room for second-order perturbative contributions, which are expected to 
improve in particular both the energy per particle $\bar E(\rho)$ and the 
spin-orbit strength $F_{so}$ \cite{holt11b}.

\section{Quasiparticle interaction in nuclear matter}

Landau's theory of normal Fermi liquids \cite{landau57} can be employed to study the bulk 
equilibrium and transport properties of strongly-interacting normal Fermi 
systems at low temperatures. The theory describes the low-energy excitations 
above the interacting ground state in terms of quasiparticles, which retain 
certain features of noninteracting particles and in a sense are weakly 
interacting. The quasiparticle interaction encodes bulk properties of the
medium and dynamical properties of the quasiparticles themselves. It can 
be computed in many-body perturbation theory by functionally differentiating
the ground-state energy twice with respect to the quasiparticle distribution
function and has the form
\begin{equation}
{\cal F}({\vec p}_1, {\vec p}_2)=\frac{1}{N_0}\sum_{L=0}^\infty \left [
F_L + F^\prime_L \vec \tau_1 \cdot
\vec \tau_2 + (G_L + G^\prime_L \vec \tau_1 \cdot \vec \tau_2) \vec \sigma_1 
\cdot \vec \sigma_2 \right ]P_L({\rm cos}\, \theta),
\label{ffunction2}
\end{equation}
where we have expanded the angular dependence in Legendre polynomials, factored 
out the density of states $N_0$ at the Fermi surface and included only the 
central components of the quasiparticle interaction.

\setlength{\tabcolsep}{.07in}
\begin{table}
\centering
\begin{tabular}{|c||c|c|c|c||c|c|c|c||c|c|c|c|} \hline
 & $F_0$ & $G_0$ & $F_0^\prime$ & $G_0^\prime$ & $F_1$ & $G_1$ &
$F^\prime_1$ & $G^\prime_1$ & $M^*/M_N$ & ${\cal K}$ [MeV] & $\beta$ [MeV] & $\delta g_l$ \\ 
\hline  \hline
$V_{\rm N3LO}^{2N}$ & $-$1.64 & 0.35 & 1.39 & 1.59 & $-$0.13 & 0.50 & 0.58 & 0.47 
               & 0.96 & $-$150 & 31 & 0.12     \\ \hline
$V_{\rm low-k}^{2N}$ & $-$1.98 & 0.58 & 1.94 & 2.14 & 0.38 & 0.83 & 0.87 & 0.80 
               & 1.13 & $-$190 & 32 & 0.07 \\ \hline \hline
$V_{\rm N3LO}^{3N}$ & $-$0.15 & 0.35 & 1.36 & 1.19 & $-$0.22 & 0.21 & 0.28 & 0.24 
               & 0.93 & 200 & 31 & 0.09     \\ \hline
$V_{\rm low-k}^{3N}$ & 1.48 & 0.22 & 1.45 & 1.48 & 0.08 & 0.37 & 0.41 & 0.39 
               & 1.03 & 530 & 29 & 0.05 \\ \hline
\end{tabular}
\caption{Dimensionless Fermi liquid parameters at nuclear matter saturation density (corresponding
to $k_F = 1.33$\,fm$^{-1}$) obtained from chiral two- and three-nucleon interactions as
described in the text.}
\label{finaltable}
\end{table}

Recently, we have employed chiral two- and three-body forces 
in a systematic study of the quasiparticle interaction in isospin-symmetric nuclear
matter \cite{holt11a,holt11c}. Already a number of observables (see eq.\ (\ref{flo}) 
for the relation to Fermi liquid parameters) were well described 
in the second-order perturbative calculation that included two-nucleon forces only \cite{holt11a}. 
In particular, the values of the quasiparticle effective mass $M^*$, the
isospin-asymmetry energy $\beta$ and the spin-isospin response parameter
$g_{NN}^\prime$ were found to be $M^*/M_N \simeq 1.0-1.1$, $\beta \simeq 
31-32$\,MeV and $g_{NN}^\prime \simeq 0.65-0.75$, 
all within empirical ranges. However, both the compression modulus ${\cal K} \simeq 
-200$\,MeV and the anomalous orbital $g$-factor $\delta g_l \simeq 0.1$ were found to
differ appreciably from the empirical values ${\cal K} \simeq 200-300$\,MeV and
$\delta g_l \simeq 0.20-0.26$ inferred from studies of giant resonances
in heavy nuclei.

As a first step toward the consistent inclusion of the N$^2$LO chiral three-nucleon
force, we have computed the first-order perturbative contribution to the $L=0,1$ Landau 
parameters of the quasiparticle interaction. Two sets of low-energy constants, given in eqs.\ (\ref{lecn3lo})--(\ref{lecvlk}),
were employed, which allowed for an estimate of the theoretical uncertainty at this order
in the calculation. The diagrammatic contributions to the quasiparticle interaction are shown
in Fig.\ \ref{ddnnint}, but since the two quasiparticles lie on the Fermi surface $|\vec p_1| = 
|\vec p_2| = k_F$, the kinematics in the present case are different than those considered 
in Section \ref{imenni}. We show in Table \ref{finaltable} 
the results from the second-order calculation including two-nucleon forces only (labeled
``$2N$'') and the results obtained by including as well the one-loop contributions from
the N$^2$LO chiral three-body force (labeled ``$3N$''). We notice that the 
most visible effect is a dramatic increase in the isotropic spin- and 
isospin-independent parameter $F_0$, which was large and negative (leading
to a negative compression modulus ${\cal K}$) when computed with only
chiral and low-momentum NN interactions, but which now attains values in 
reasonable agreement with empirical constraints from giant monopole resonances. 
In comparison, the contributions to the other Landau parameters from the chiral 
three-nucleon force considered here are much smaller, and most observables
 remain in good agreement with their empirical values:
\begin{eqnarray}
{\rm Effective \, \, mass:} \hspace{.2in} \frac{M^*}{M_N} = 1+F_1/3 =& 0.98 \pm 0.05, &[0.7-1.0]
\nonumber \\
{\rm Anomalous \, \, orbital \, \,} g\mbox{-factor:} \hspace{.2in} \delta 
g_l = \frac{F^\prime_1-F_1}{6(1+F_1/3)}  =& 0.07 \pm 0.02, & [0.20-0.26] \nonumber \\
{\rm Compression \, \, modulus:} \hspace{.2in} {\cal K} =\frac{3k_F^2}{M^*} 
\left (1+F_0\right ) \, =& (370 \pm 160) \, {\rm MeV},\hspace{.2in}
& [200-300]\, {\rm MeV} \nonumber \\
{\rm Isospin\, \,  asymmetry \, \, energy:} \hspace{.2in} \beta =\frac{k_F^2}{6M^*}(1+F_0^\prime)=& 
(30 \pm 1) \, {\rm MeV}, & [30-36]\, {\rm MeV} \nonumber \\
{\rm Spin\mbox{-}isospin\, \, response:} \hspace{.2in} g_{NN}^\prime =\frac{4M_N^2}{g_{\pi N}^2N_0} G_0^\prime
=& 0.55 \pm 0.03\, &[0.6-0.7].
\label{flo}
\end{eqnarray}
In these equations, $g_{\pi N} \simeq 13.2$ is the strong $\pi N$ coupling constant and the
values in brackets represent empirical estimates (for further details, see ref.\ \cite{holt11c}).
Moreover, the theoretical 
uncertainty resulting from different choices of cutoff scale and low-energy
constants is reduced with the inclusion of three-nucleon forces. The one observable
which remains in disagreement with its empirical value is 
the anomalous orbital $g$-factor $\delta g_l$, which lies well below the 
value $\delta g_l = 0.20 - 0.26$ extracted from a sum-rule analysis of giant 
dipole resonances. As suggested in ref.\ \cite{holt11c}, this could be 
remedied if higher-order perturbative calculations reduce even moderately the quasiparticle
effective mass $M^*$. Finally, we notice that all $L=1$ Landau parameters decrease
with the addition of three-nucleon forces, which results in an effective 
interaction of apparent short range.

\section{Conclusion}
 
In this talk we have focused on our recent efforts to incorporate the 
leading-order chiral three-nucleon force in many-body perturbation theory
calculations of dense nuclear systems. This
work has set the foundation for future shell model and mean field theory
calculations of finite nuclei, and our investigations of infinite 
isospin-symmetric nuclear 
matter can be broadened to include pure neutron matter as a first 
approximation to neutron-star matter. In the future we look forward to employing
the recently completed subleading chiral three-nucleon 
force \cite{bernard08}, which will allow for consistent calculations
up to order N$^3$LO in the chiral expansion.

\end{document}